\documentclass[conference]{IEEEtran}
\IEEEoverridecommandlockouts
\usepackage{cite}
\usepackage{amsmath,amssymb,amsfonts}
\usepackage{algorithmic}
\usepackage{graphicx}
\usepackage{polski}
\usepackage{textcomp}
\usepackage{xcolor}
\def\BibTeX{{\rm B\kern-.05em{\sc i\kern-.025em b}\kern-.08em
    T\kern-.1667em\lower.7ex\hbox{E}\kern-.125emX}}
\begin{document}

\title{Evolve the Model Universe of a System Universe

\thanks{This work is supported by the Co-tester project (No. 314544) funded by the Research Council of Norway. }
}

\author{\IEEEauthorblockN{1\textsuperscript{st} Tao Yue}
\IEEEauthorblockA{\textit{Simula Research Laboratory} \\
Oslo, Norway \\
tao@simula.no, 0000-0003-3262-5577}
\and
\IEEEauthorblockN{2\textsuperscript{nd} Shaukat Ali}
\IEEEauthorblockA{\textit{Simula Research Laboratory} \\
\textit{Oslo Metropolitan University}\\
Oslo, Norway \\
shaukat@simula.no, 0000-0002-9979-3519}
}

\maketitle

\begin{abstract}
Uncertain, unpredictable, real-time, and lifelong evolution causes operational failures in intelligent software systems, leading to significant damages, safety and security hazards, and tragedies. To fully unleash such systems’ potential and facilitate their wider adoption, ensuring the trustworthiness of their decision-making under uncertainty is the prime challenge. To overcome this challenge, an intelligent software system and its operating environment should be continuously monitored, tested, and refined during its lifetime operation. Existing technologies, such as digital twins, can enable continuous synchronisation with such systems to reflect their most up-to-date states. Such representations are often in the form of prior-knowledge-based and machine-learning models, together called ‘model universe’.
In this paper, we present our vision of combining techniques from software engineering, evolutionary computation, and machine learning to support the model universe evolution. 

\end{abstract}


\section{Motivation}
Intelligent software systems are transforming business, life, and the global economy. Machine learning (ML) techniques are often employed in such systems to enable nontrivial autonomous decision-making under uncertainties, thereby being intelligent \cite{lwakatare2020large}. Such systems are prone to unforeseen situations in operation due to several factors, including 1) various degrees of uncertainty in physical environments and networks; 2) the probabilistic, non-backwards-traceable nature of the inner workings of the ML techniques employed; 3) unpredictable or design-time-unknown operating environments; and 4) the systems’ own continuous and lifelong learning/evolution. 

Such uncertain, unpredictable, real-time, and lifelong evolution causes operational failures in intelligent software systems, leading to significant damages, safety and security hazards, and tragedies. Hence, ensuring the dependability of such systems at design and development time alone is insufficient to ensure their dependability in real-world operation. Current approaches (e.g., testing) are insufficient before such systems are deployed since it is impossible to know all the critical situations these systems will experience in the real world. Some of these situations appear only during their operations, and it is also hard (if even possible) to predict when and why. \textit{To fully unleash intelligent software systems’ potential and facilitate their wider adoption, ensuring the trustworthiness of their decision-making under uncertainty is the prime challenge.}

Hence, intelligent software systems should be continuously monitored, tested, and refined with real-world data to ensure they can gracefully handle all uncertain and unknown situations during their lifetime. Current technologies, such as digital twins (digital and live representations of systems), can enable continuous synchronisation with the systems to reflect their most up-to-date states \cite{10.1007/978-3-030-83723-5_5}. Such representations are often in the form of prior-knowledge-based and ML models (i.e., model universe). The former is widely used to represent software systems; however, such models have a limited capability to support the runtime analyses, reasoning, validation, and validation of intelligent software systems during their operations in uncertain environments. This is simply because the prior knowledge required to create these models is only partially available and, in some cases, has yet to be discovered. Even worse, soon after their creation, these models become obsolete and useless. This obsolescence is accelerated when ML techniques are employed since ML models face performance degradation over time due to, for example, data drift, and they must inevitably evolve when more data becomes available during the operation of such systems. \textit{To stay alive and, therefore, valid and functional, the model universe must continuously evolve to faithfully represent the system of interest and its environment (i.e., system universe)}. 


\section{Concept Formulation and State-of-the-art}
\begin{figure}[!tb]
\centering
\includegraphics[width=0.8\columnwidth]{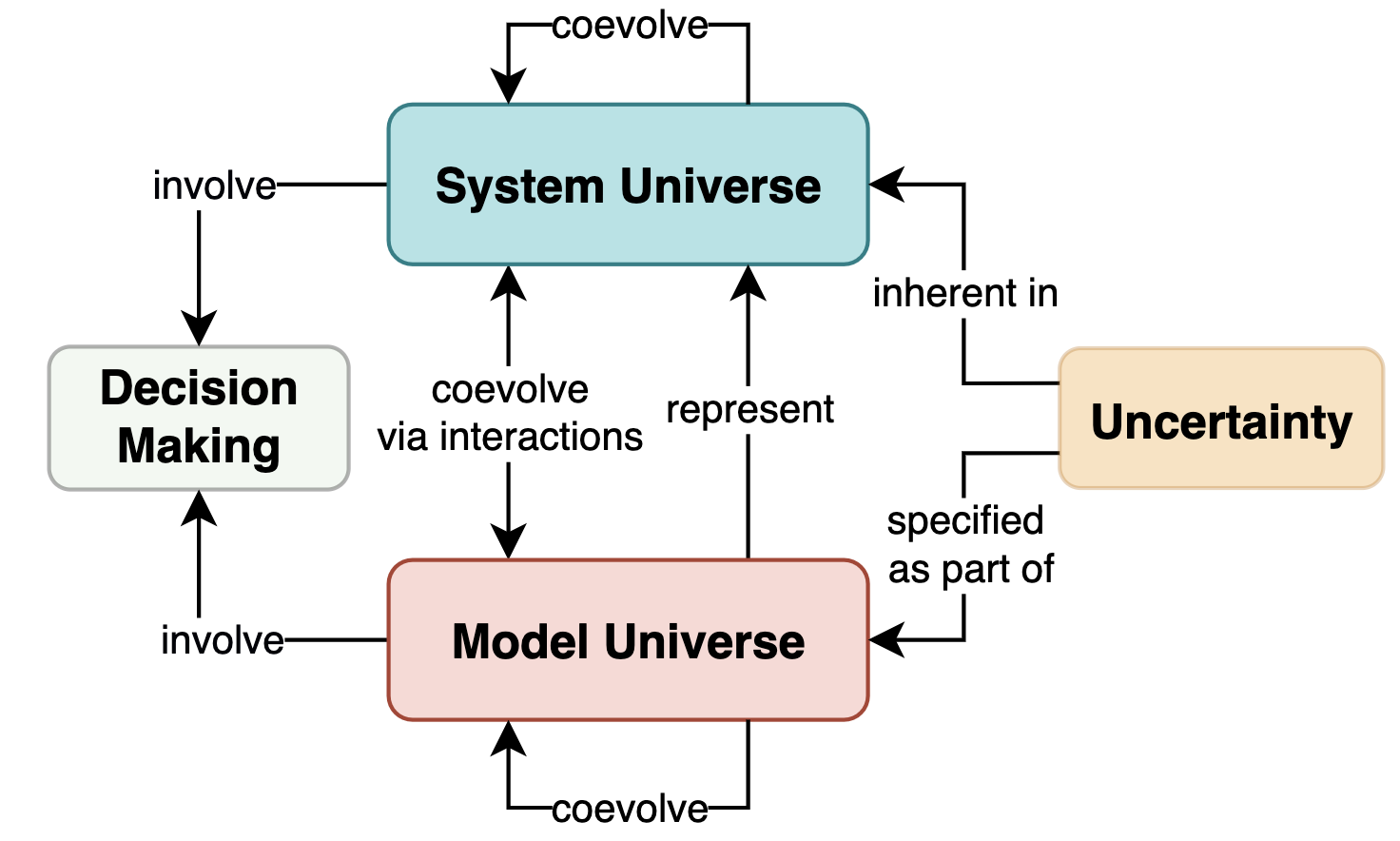}
\caption{Concepts and their relations. A \textit{system universe} (an intelligent software system and its operating environment) is represented by its \textit{model universe} (prior-knowledge-based and ML models with intelligent capabilities). }
\label{fig:overview}
\end{figure}

We present the key concepts and their relationships in Figure \ref{fig:overview}. In the rest of the section, we discuss them in detail.
\subsubsection{Model and system universes}
A model is considered an ‘informative representation of an object, person or system’~\cite{Model}. System models are simplified representations of reality’s essential or relevant entities and their properties at particular points in time and/or space that form particular interests, importance, and concerns and serve specific purposes. Models are prevailingly used in software/system engineering and are often classified into two categories regarding their construction: prior-knowledge-based and data-driven models. The former comprises models such as 3D models created with simulators (e.g., for virtual surgical planning), Simulink, and Systems Modeling Language (SysML) models for model-based system engineering \cite{5722047,WANG2023103804}; the latter mainly refers to ML models, e.g., AlexNet for image classification \cite{krizhevsky2017imagenet} and YOLO for object detection \cite{redmon2017yolo9000}. 

The model universe of a system universe provides the proper basis for reasoning about the system universe and enables decision makings of all kinds. Here, we use the term ‘universe’ to emphasise that our universe is 95\% unknown \cite{aghanim2021planck}; similarly, the system and model universes contain many unknowns. Furthermore, knowing the universe is about understanding its formation and evolution, and such an understanding is essential for building a theory based on which scientific extrapolations can be made about the future of the universe. 

\subsubsection{Uncertainties}
The concept of ‘uncertainty’ can be traced back to the philosophical question about the certainty of knowledge, debated by the ancient Greek philosophers, including Aristotle. Uncertainty has attracted significant attention since it is inherent in intelligent software systems and their operating environments \cite{zhang2020uncertainty1, shin2021uncertainty, catak2022uncertainty,10.1145/3540250.3558955,10.1145/3576041}. One representative example is the uncertain operating environments of autonomous driving vehicles (i.e., external uncertainties) and inherent uncertainties of their behaviours due to the use of ML models for perception and path planning, among other decision-making tasks (i.e., internal uncertainties). Moreover, models aimed at understanding, reasoning, and predicting system behaviours make assumptions of all kinds. Therefore, a model universe devised for a system universe contains two types of uncertainties: \textit{objective uncertainties}, which refer to phenomena whose existence and nature are independent of any observing agency, and \textit{subjective uncertainties}, which refer to information existing within some agency derived from that agency’s observations and/or reasoning (i.e., belief agents) \cite{10.1007/978-3-319-42061-5_16}. When gaining more knowledge, subjective uncertainties can evolve into objective ones. 

Uncertainties can also be classified into \textit{shallow uncertainties} and \textit{deep uncertainties}. Being shallow means that the probabilities of the outcomes are well known; therefore, future events can be reasonably predicted by the past. On the other hand, deep uncertainties refer to contexts in which the probabilities of the outcomes are poorly known, unknown, or unknowable, such that past events can give little insight into future ones \cite{aghanim2021planck}. Though deep uncertainties have been discussed in, for example, natural hazard risk assessment \cite{almeida2017dealing} and financial investments in climate change \cite{narita2022evaluating}, they are rarely recognised in software engineering. 

Uncertainties in ML refer to the lack of confidence in an ML model’s output. Estimating them is essential to determine if they are low enough that the output can be trusted. Typically, they are classified into (irreducible) \textit{aleatory uncertainties} and (reducible) \textit{epistemic uncertainties}, referring to the inherent stochasticity of the observations and the lack of training data. The software engineering community has only recently studied uncertainty in ML. It, therefore, still primarily focuses on, for example, applying uncertainty quantification methods to supervise ML systems \cite{weiss2022uncertainty, xu2022uncertainty, 10.1145/3582571, 9564336} with tool support \cite{weiss2021uncertainty, zou2022neuraluq, chung2021uncertainty}. 
All these works limit the scope to uncertainties caused by the (inherent) limitations of learned ML models, that is, not covering \textit{data quality uncertainty} (e.g., where the quality of the input data is lower than the training data’s) and \textit{scope compliance uncertainty} (concerning differences between a modelled context and its intended application context), as classified in \cite{klas2018uncertainty}.
The gap between uncertainties understood and captured in prior-knowledge-based models of the model universe and uncertainties recognised and quantified in its ML models is also not yet bridged. 

\subsubsection{Evolution of Prior-knowledge-based and ML models}
The common practice of software engineering is to manually and offline evolve models created based on prior knowledge with methods such as inference engines (e.g., Daikon \cite{ernst2007daikon}). For instance, Zhang et al. proposed data-augmented model evolution methods supported by model execution and simulation techniques \cite{zhang2017uncertainty} to semi-automatically evolve Unified Modeling Language (UML) state machines and uncertainty measurements. Considering the multi-paradigm modelling (MPM) nature of the model universe, its evolution involves the evolution of models belonging to the same modelling paradigm (e.g., SysML block definition diagrams and state machines) and the coevolution of models across MPM (e.g., Modelica models and 3D CAD models). In the literature, solutions have been proposed for MPM and co-simulations \cite{de2020simulation, barivsic2022multi}, but not for the coevolution of such models. For instance, UncerTolve \cite{ZHANG2017140} advanced state of the art by using real operational data from CPSs to evolve test models in UML and subjective uncertainties offline, and DeepCollision \cite{9712397} and LiveTCM \cite{9592382} evolve test scenarios of autonomous driving vehicles in a 3D virtual environment with reinforcement learning. 

ML models are often statically learned from historical data with ML techniques. Most domain adaptation and lifelong learning methods address data drift offline, requiring the availability of both source and target domain information beforehand, an assumption that prevents them from being applied in real contexts. For instance, RISE-DT \cite{xu2022uncertainty} is one such approach: it automatically evolves a digital twin (with its model captured in automata and its capability enabled with an ML model) to be applied to a different application context of industrial elevator systems with transfer learning offline. However, online methods must fit the real-time context of evolving the model universe. Online domain adaptation has recently been proposed to continuously handle data drift for semantic segmentation under ever-changing conditions during deployment \cite{panagiotakopoulos2022online}. In addition, to learn from non-stationary (where data become incrementally available over time) real-world data, lifelong or continual learning – that is, continually evolving (via acquiring, turning and transferring) knowledge throughout lifespans across domains – seems promising \cite{parisi2019continual}. The most recent advance in this field is online lifelong and continual learning \cite{mai2022online, liu2020learning}, suitable for model universe evolution. Still, we need to see applications and empirical studies, which are currently largely unavailable. Despite these efforts, there needs to be a solution enabling the holistic coevolution of the model universe.

\subsubsection{Coevolution of model and system universes}
Due to uncertainties, a system universe is naturally an evolutionary reality. Therefore, its corresponding model universe should be an evolutionary model of the evolutionary reality. Any mechanical model (without the dynamics of changing for ‘good’) is doomed to be useless. As well put by George Box in \cite{BOX1979201}, ‘all models are wrong, but some are useful’. This statement initially referred to statistical models but now generally applies to all models. During modelling, assumptions are made to understand and predict the system universe. When the system universe evolves, to be valid, the model universe needs to evolve itself accordingly by 1) validating knowns (captured in the model universe) with new information obtained from the system universe, 2) refining subjective uncertainties, 3) discovering unknowns to invalidate captured assumptions, and 4) recognise the unknowable.

Without a suitable evolution mechanism, the difference between the model and system universes becomes prominent, and decisions based on an outdated model universe are prone to errors. To maintain the model universe’s usefulness, it must be continuously evolved to remain \textit{alive}. We, therefore, define \textit{evolution} in the model universe as its progression towards a direction from an uncertain or worse state to a more certain or better one in terms of supporting the system universe’s development, operation, and maintenance. 
We consider \textit{coevolution} in universes as an evolution involving interactions of more than one model type in the model universe or interactions across the model and system universes.  

\subsubsection{Coevolutionary algorithms}
Evolutionary computation has been applied to solve many optimisation problems, e.g., test optimisation \cite{arrieta2019search, panichella2018large, silva2017systematic}, product configuration optimisation \cite{henard2015combining, lopez2015systematic}, and optimisations in requirements engineering \cite{zhang2020uncertainty, 10.1007/978-3-540-69062-7_8}. A subset of evolutionary algorithms, coevolutionary algorithms evaluate an individual’s fitness based on how the individual performs against others in the population \cite{xu2006low}, known as indirect fitness. Relationships such as competition and cooperation among individuals are vital in designing coevolutionary algorithms, which can help simulate real-world scenarios such as pedestrian detection \cite{xu2006low} and search for game-playing strategies \cite{rosin1997new}. 

Coevolutionary algorithms are often implemented in different metaheuristic algorithms – such as genetic algorithms (GA), genetic programming, and differential evolution \cite{8454482} – evidence shows that coevolutionary and conventional evolutionary algorithms can complement each other very well. In software engineering, although coevolution has been leveraged in addressing software development and testing challenges – as in test case generation by coevolving test inputs and test oracles \cite{ouni2013maintainability}, automatic programming \cite{arcuri2014co} and software correction \cite{wilkerson2012multi}, and the coevolution of models and tests \cite{koos2009automatic} – it is still largely unexplored for addressing complex and practical problems requiring novel coevolution strategies implemented in suitable evolutionary algorithms. 

\subsubsection{Epigenetics and epigenetic algorithms} 
‘Epigenetics studies heritable changes in gene expression that occur without changes in DNA sequence’ in response to environmental changes \cite{bollati2010environmental}. An example is an octopus temporarily changing colour to one not encoded in its DNA in response to environmental threats \cite{liscovitch2017trade}. Several recent reviews have studied the role of epigenetics in domesticated animals \cite{vogt2017facilitation}, plants \cite{kelly2015churchill}, and humans \cite{kanherkar2014epigenetics}. For instance, diverse environmental behaviours (e.g. stress, exercise, and exposure to a toxic environment) bring epigenetic changes (both positive and adverse) in humans during their life spans.

We postulate that epigenetic algorithms fit model universe evolution well because 1) genes passed down from parents (genetic inheritance) cannot react to sudden changes in the environment by themselves, but epigenetic inheritance, by controlling how genes work, allows for fast adaptation when appropriate, increasing the speed of convergence while maintaining stability in a changing environment; 2) epigenetic mechanisms have the potential to respond – and probably in most cases positively – to all types of uncertainties if we constrain the direction of evolution; and 3) epigenetic and coevolutionary algorithms are both nature-inspired and hence can be naturally incorporated. 

Only a few epigenetic algorithms have been proposed: epiGA \cite{stolfi2018epigenetic} implements gene silencing and integrates it into GA to control how genes are expressed or turned on or off in response to environmental uncertainties; EpiLearn \cite{mukhlish2020reward} encodes dynamic environmental changes as an epigenetic layer in a learning process to allow for adaptive and efficient learning; and RELEpi \cite{mukhlish2020reward} supports the coevolving decision making of groups of agents (swarms) in uncertain environments, although it has not yet proven effective for real-world problems. 

Though these works demonstrate that epigenetic algorithms are a promising direction for coping with uncertainty and unknowns, we are far from being able to apply them to handle uncertainties in model universe evolution due to 1) rarely seeing implementations of epigenetic mechanisms from biology, 2) the lack of real-world applications, and 3) the unavailability of experimental frameworks, empirical data, and research communities. This is because epigenetics in biology is relatively new and complex, and epigenetic encoding for real applications requires a deep understanding of application contexts, problems to be solved, and epigenetic mechanisms.   

\textit{Concluding remark.}
There is no \textit{holistic} method for evolving the model universe of a system universe under unknown uncertainties because the model universe’s evolution: 1) needs to be online and autonomous, 2) is triggered at different times and in different spaces, 3) is tightly entangled with uncertainties of various types and degrees, 4) needs to coordinate diverse modelling paradigms, and 5) needs to interact with the system universe during its operation efficiently. 

\section{The Way forward}

\subsubsection{Identify universe coevolution patterns} 
We first need to understand primary coevolution triggers (e.g., discovering unknowns in prior-knowledge–based models, obtaining new data for adapting ML models), the time points that trigger each coevolution (e.g., upon receiving a new batch of data, as soon as an uncertain event occurs), and the conditions under which each model needs to be evolved to keep themselves alive and the data requirements (e.g., quality and quantity). Following the common practice of model-based engineering, these understandings can be specified as a metamodel, based on which methodologies and tools to specify, characterise, and automatically identify each pattern can be proposed.

\subsubsection{Inspired by coevolution mechanisms in nature}
Coevolution mechanisms in nature can provide inspiration for designing and applying coevolutionary algorithms; however, not all coevolution mechanisms can be directly implemented in coevolutionary algorithms, mainly because these mechanisms are complex and we have only a limited understanding of them. It is therefore important to only implement their essential features. For instance, some interactions between the model and system universes might be mapped to commensalism, where the model universe benefits from the interactions with the system universe (e.g., evolving itself with data received from the system universe) while the system universe remains unaffected. Future research is needed to systematically map universe coevolution patterns (representing the problems to be solved) to coevolution mechanisms in nature, which will leverage the identification of existing coevolutionary algorithms and the development of novel coevolutionary algorithms to enable model universe evolution.

\subsubsection{Develop uncertainty taxonomy, metamodel, quantification, and management methods}
We need to develop a comprehensive uncertainty taxonomy to support end-to-end uncertainty management, characterising and quantifying uncertainties (with the uncertainty metamodel to be devised) which results in uncertainty models being managed as part of the model universe. We will also need to integrate various uncertainty quantification methods for ML and prior-knowledge–based models to efficiently enable holistic end-to-end uncertainty quantification that involves more than one quantification method, for example, connecting subjective uncertainties from prior-knowledge models to objective uncertainties from sensory data and, further, to uncertainties in ML models’ predictions, which could lead to uncertain decision making and the actuation of physical devices. A solution needs to systematically select and base itself on various uncertainty-related theories, such as probability theory for quantifying the likelihood of known outcomes, possibility theory for situations in which the probability of an event is unknown or unknowable, Bayesian decision theory and Bayesian updating to update the prior knowledge of belief agents about given events, prospect theory for (subjective) human decision making under uncertainty and risk (e.g., pedestrians crossing roads), chaos theory for modelling, analysing, and improving system robustness, Dempster-Shafer theory in handling situations lacking complete information, and combinations of theories, powering end-to-end uncertainty-aware model universe evolution. 

\subsubsection{Propose coevolutionary algorithms and epigenetics-inspired algorithms}
An envisioned solution must be autonomous, data-augmented, and online, which puts high requirements on its efficiency. To achieve this, we must first rely on coevolutionary algorithms by developing optimal encoding mechanisms for each coevolution pattern, defining subjective internal measures for fitness, defining adaptive problem decomposition structures, and overcoming challenges such as avoiding local optima, scaling, and measuring performance. In response to uncertainties, we need to develop epigenetics-inspired algorithms that mimic biological adaptations in species by encoding each model universe’s evolution pattern in the form of genomes and epigenomes (both responsible for the regulation and expression of genetic information), implementing generic epigenetic operators based on three epigenetic mechanisms (DNA methylation, histone modification, and RNA editing), and simulating uncertainties in gene expressions through epigenetic changes by applying epigenetic operators to genes in the model universe through mechanisms such as the introduction of epigenetic drift (i.e., where the epigenetic marks change over time) and combing different epigenetic operators.

\subsubsection{Design multi-agent evolutionary reinforcement learning methods}
We need to introduce coevolutionary algorithms to multi-agent reinforcement learning by matching each model in the model universe as a learning agent, its environment as the system universe, and other models (i.e., agents) in the model universe. For instance, for policy-based reinforcement learning, one can let agents compete or cooperate (or other coevolution mechanisms) to learn from their interactions, evolve the agents’ policies via the coevolutionary process, and design the agents’ rewards to encourage or discourage their behaviours. We will introduce epigenetic mechanisms to coevolutionary algorithms based on uncertainties under study, for example, by controlling gene expressions in response to the environmental changes of individuals in an agent’s population, influencing the evolution of individuals by controlling genetic operators such as mutation rates, and controlling the selection of individuals for reproduction. When integrating epigenetic and coevolutionary algorithms with multi-agent reinforcement learning, we can use coevolutionary algorithms to evolve the policies of multiple agents in a reinforcement learning setting and use the epigenetic mechanisms to modulate the agents’ policies based on their experiences and interactions with other agents/models and the environment (i.e., the system universe).

\textit{Impact.} Intelligent software systems are used in many applications, such as healthcare, agriculture, transportation, and manufacturing, and have an enormous impact on our lives and demand a high degree of dependability on these systems’ operations. With the envisioned solution, the dependability of the current and future intelligent software systems will be significantly improved through fully-fledged model universes capable of robustly dealing with uncertainties in real time.

\bibliographystyle{IEEEtran}
\bibliography{bib}

\begin{thebibliography}{10}
\providecommand{\url}[1]{#1}
\csname url@samestyle\endcsname
\providecommand{\newblock}{\relax}
\providecommand{\bibinfo}[2]{#2}
\providecommand{\BIBentrySTDinterwordspacing}{\spaceskip=0pt\relax}
\providecommand{\BIBentryALTinterwordstretchfactor}{4}
\providecommand{\BIBentryALTinterwordspacing}{\spaceskip=\fontdimen2\font plus
\BIBentryALTinterwordstretchfactor\fontdimen3\font minus
  \fontdimen4\font\relax}
\providecommand{\BIBforeignlanguage}[2]{{%
\expandafter\ifx\csname l@#1\endcsname\relax
\typeout{** WARNING: IEEEtran.bst: No hyphenation pattern has been}%
\typeout{** loaded for the language `#1'. Using the pattern for}%
\typeout{** the default language instead.}%
\else
\language=\csname l@#1\endcsname
\fi
#2}}
\providecommand{\BIBdecl}{\relax}
\BIBdecl

\bibitem{lwakatare2020large}
L.~E. Lwakatare, A.~Raj, I.~Crnkovic, J.~Bosch, and H.~H. Olsson, ``Large-scale
  machine learning systems in real-world industrial settings: A review of
  challenges and solutions,'' \emph{Information and software technology}, vol.
  127, p. 106368, 2020.

\bibitem{10.1007/978-3-030-83723-5_5}
T.~Yue, P.~Arcaini, and S.~Ali, ``Understanding digital twins for
  cyber-physical systems: A conceptual model,'' in \emph{Leveraging
  Applications of Formal Methods, Verification and Validation: Tools and
  Trends}, T.~Margaria and B.~Steffen, Eds.\hskip 1em plus 0.5em minus
  0.4em\relax Cham: Springer International Publishing, 2021, pp. 54--71.

\bibitem{Model}
\BIBentryALTinterwordspacing
``Model,'' 2023. [Online]. Available: \url{https://en.wikipedia.org/wiki/Model}
\BIBentrySTDinterwordspacing

\bibitem{5722047}
A.~L. Ramos, J.~V. Ferreira, and J.~Barceló, ``Model-based systems
  engineering: An emerging approach for modern systems,'' \emph{IEEE
  Transactions on Systems, Man, and Cybernetics, Part C (Applications and
  Reviews)}, vol.~42, no.~1, pp. 101--111, 2012.

\bibitem{WANG2023103804}
\BIBentryALTinterwordspacing
T.~Wang, C.~Tan, L.~Huang, Y.~Shi, T.~Yue, and Z.~Huang, ``Simplexity testbed:
  A model-based digital twin testbed,'' \emph{Computers in Industry}, vol. 145,
  p. 103804, 2023. [Online]. Available:
  \url{https://www.sciencedirect.com/science/article/pii/S0166361522002007}
\BIBentrySTDinterwordspacing

\bibitem{krizhevsky2017imagenet}
A.~Krizhevsky, I.~Sutskever, and G.~E. Hinton, ``Imagenet classification with
  deep convolutional neural networks,'' \emph{Communications of the ACM},
  vol.~60, no.~6, pp. 84--90, 2017.

\bibitem{redmon2017yolo9000}
J.~Redmon and A.~Farhadi, ``Yolo9000: better, faster, stronger,'' in
  \emph{Proceedings of the IEEE conference on computer vision and pattern
  recognition}, 2017, pp. 7263--7271.

\bibitem{aghanim2021planck}
N.~Aghanim, Y.~Akrami, M.~Ashdown, J.~Aumont, C.~Baccigalupi, M.~Ballardini,
  A.~Banday, R.~Barreiro, N.~Bartolo, S.~Basak \emph{et~al.}, ``Planck 2018
  results-vi. cosmological parameters (corrigendum),'' \emph{Astronomy \&
  Astrophysics}, vol. 652, p.~C4, 2021.

\bibitem{zhang2020uncertainty1}
X.~Zhang, ``Uncertainty-guided testing and robustness enhancement for deep
  learning systems,'' in \emph{Proceedings of the ACM/IEEE 42nd International
  Conference on Software Engineering: Companion Proceedings}, 2020, pp.
  101--103.

\bibitem{shin2021uncertainty}
S.~Y. Shin, K.~Chaouch, S.~Nejati, M.~Sabetzadeh, L.~C. Briand, and F.~Zimmer,
  ``Uncertainty-aware specification and analysis for hardware-in-the-loop
  testing of cyber-physical systems,'' \emph{Journal of Systems and Software},
  vol. 171, p. 110813, 2021.

\bibitem{catak2022uncertainty}
F.~O. Catak, T.~Yue, and S.~Ali, ``Uncertainty-aware prediction validator in
  deep learning models for cyber-physical system data,'' \emph{ACM Transactions
  on Software Engineering and Methodology (TOSEM)}, vol.~31, no.~4, pp. 1--31,
  2022.

\bibitem{10.1145/3540250.3558955}
\BIBentryALTinterwordspacing
L.~Han, T.~Yue, S.~Ali, A.~Arrieta, and M.~Arratibel, ``Are elevator software
  robust against uncertainties? results and experiences from an industrial case
  study,'' ser. ESEC/FSE 2022.\hskip 1em plus 0.5em minus 0.4em\relax New York,
  NY, USA: Association for Computing Machinery, 2022, p. 1331–1342. [Online].
  Available: \url{https://doi.org/10.1145/3540250.3558955}
\BIBentrySTDinterwordspacing

\bibitem{10.1145/3576041}
\BIBentryALTinterwordspacing
L.~Han, S.~Ali, T.~Yue, A.~Arrieta, and M.~Arratibel, ``Uncertainty-aware
  robustness assessment of industrial elevator systems,'' vol.~32, no.~4, may
  2023. [Online]. Available: \url{https://doi.org/10.1145/3576041}
\BIBentrySTDinterwordspacing

\bibitem{10.1007/978-3-319-42061-5_16}
M.~Zhang, B.~Selic, S.~Ali, T.~Yue, O.~Okariz, and R.~Norgren, ``Understanding
  uncertainty in cyber-physical systems: A conceptual model,'' in
  \emph{Modelling Foundations and Applications}, A.~W{\k{a}}sowski and
  H.~L{\"o}nn, Eds.\hskip 1em plus 0.5em minus 0.4em\relax Cham: Springer
  International Publishing, 2016, pp. 247--264.

\bibitem{almeida2017dealing}
S.~Almeida, E.~A. Holcombe, F.~Pianosi, and T.~Wagener, ``Dealing with deep
  uncertainties in landslide modelling for disaster risk reduction under
  climate change,'' \emph{Natural Hazards and Earth System Sciences}, vol.~17,
  no.~2, pp. 225--241, 2017.

\bibitem{narita2022evaluating}
D.~Narita, I.~Sato, D.~Ogawada, and A.~Matsumura, ``Evaluating the robustness
  of project performance under deep uncertainty of climate change: A case study
  of irrigation development in kenya,'' \emph{Climate Risk Management},
  vol.~36, p. 100426, 2022.

\bibitem{weiss2022uncertainty}
M.~Weiss and P.~Tonella, ``Uncertainty quantification for deep neural networks:
  An empirical comparison and usage guidelines,'' \emph{arXiv preprint
  arXiv:2212.07118}, 2022.

\bibitem{xu2022uncertainty}
Q.~Xu, S.~Ali, T.~Yue, and M.~Arratibel, ``Uncertainty-aware transfer learning
  to evolve digital twins for industrial elevators,'' in \emph{Proceedings of
  the 30th ACM Joint European Software Engineering Conference and Symposium on
  the Foundations of Software Engineering}, 2022, pp. 1257--1268.

\bibitem{10.1145/3582571}
\BIBentryALTinterwordspacing
Q.~Xu, S.~Ali, and T.~Yue, ``Digital twin-based anomaly detection with
  curriculum learning in cyber-physical systems,'' \emph{ACM Trans. Softw. Eng.
  Methodol.}, feb 2023, just Accepted. [Online]. Available:
  \url{https://doi.org/10.1145/3582571}
\BIBentrySTDinterwordspacing

\bibitem{9564336}
F.~O. Catak, T.~Yue, and S.~Ali, ``Prediction surface uncertainty
  quantification in object detection models for autonomous driving,'' in
  \emph{2021 IEEE International Conference on Artificial Intelligence Testing
  (AITest)}, 2021, pp. 93--100.

\bibitem{weiss2021uncertainty}
M.~Weiss and P.~Tonella, ``Uncertainty-wizard: Fast and user-friendly neural
  network uncertainty quantification,'' in \emph{2021 14th IEEE Conference on
  Software Testing, Verification and Validation (ICST)}.\hskip 1em plus 0.5em
  minus 0.4em\relax IEEE, 2021, pp. 436--441.

\bibitem{zou2022neuraluq}
Z.~Zou, X.~Meng, A.~F. Psaros, and G.~E. Karniadakis, ``Neuraluq: A
  comprehensive library for uncertainty quantification in neural differential
  equations and operators,'' \emph{arXiv preprint arXiv:2208.11866}, 2022.

\bibitem{chung2021uncertainty}
Y.~Chung, I.~Char, H.~Guo, J.~Schneider, and W.~Neiswanger, ``Uncertainty
  toolbox: an open-source library for assessing, visualizing, and improving
  uncertainty quantification,'' \emph{arXiv preprint arXiv:2109.10254}, 2021.

\bibitem{klas2018uncertainty}
M.~Kl{\"a}s and A.~M. Vollmer, ``Uncertainty in machine learning applications:
  A practice-driven classification of uncertainty,'' in \emph{Computer Safety,
  Reliability, and Security: SAFECOMP 2018 Workshops, ASSURE, DECSoS, SASSUR,
  STRIVE, and WAISE, V{\"a}ster{\aa}s, Sweden, September 18, 2018, Proceedings
  37}.\hskip 1em plus 0.5em minus 0.4em\relax Springer, 2018, pp. 431--438.

\bibitem{ernst2007daikon}
M.~D. Ernst, J.~H. Perkins, P.~J. Guo, S.~McCamant, C.~Pacheco, M.~S. Tschantz,
  and C.~Xiao, ``The daikon system for dynamic detection of likely
  invariants,'' \emph{Science of computer programming}, vol.~69, no. 1-3, pp.
  35--45, 2007.

\bibitem{zhang2017uncertainty}
M.~Zhang, S.~Ali, T.~Yue, and R.~Norgre, ``Uncertainty-wise evolution of test
  ready models,'' \emph{Information and Software Technology}, vol.~87, pp.
  140--159, 2017.

\bibitem{de2020simulation}
W.~de~Paula~Ferreira, F.~Armellini, and L.~A. De~Santa-Eulalia, ``Simulation in
  industry 4.0: A state-of-the-art review,'' \emph{Computers \& Industrial
  Engineering}, vol. 149, p. 106868, 2020.

\bibitem{barivsic2022multi}
A.~Bari{\v{s}}i{\'c}, I.~Ruchkin, D.~Savi{\'c}, M.~A. Mohamed, R.~Al-Ali, L.~W.
  Li, H.~Mkaouar, R.~Eslampanah, M.~Challenger, D.~Blouin \emph{et~al.},
  ``Multi-paradigm modeling for cyber--physical systems: A systematic mapping
  review,'' \emph{Journal of Systems and Software}, vol. 183, p. 111081, 2022.

\bibitem{ZHANG2017140}
\BIBentryALTinterwordspacing
M.~Zhang, S.~Ali, T.~Yue, and R.~Norgre, ``Uncertainty-wise evolution of test
  ready models,'' \emph{Information and Software Technology}, vol.~87, pp.
  140--159, 2017. [Online]. Available:
  \url{https://www.sciencedirect.com/science/article/pii/S0950584917302161}
\BIBentrySTDinterwordspacing

\bibitem{9712397}
C.~Lu, Y.~Shi, H.~Zhang, M.~Zhang, T.~Wang, T.~Yue, and S.~Ali, ``Learning
  configurations of operating environment of autonomous vehicles to maximize
  their collisions,'' \emph{IEEE Transactions on Software Engineering},
  vol.~49, no.~1, pp. 384--402, 2023.

\bibitem{9592382}
Y.~Shi, C.~Lu, M.~Zhang, H.~Zhang, T.~Yue, and S.~Ali, ``Restricted natural
  language and model-based adaptive test generation for autonomous driving,''
  in \emph{2021 ACM/IEEE 24th International Conference on Model Driven
  Engineering Languages and Systems (MODELS)}, 2021, pp. 101--111.

\bibitem{panagiotakopoulos2022online}
T.~Panagiotakopoulos, P.~L. Dovesi, L.~H{\"a}renstam-Nielsen, and M.~Poggi,
  ``Online domain adaptation for semantic segmentation in ever-changing
  conditions,'' in \emph{Computer Vision--ECCV 2022: 17th European Conference,
  Tel Aviv, Israel, October 23--27, 2022, Proceedings, Part XXXIV}.\hskip 1em
  plus 0.5em minus 0.4em\relax Springer, 2022, pp. 128--146.

\bibitem{parisi2019continual}
G.~I. Parisi, R.~Kemker, J.~L. Part, C.~Kanan, and S.~Wermter, ``Continual
  lifelong learning with neural networks: A review,'' \emph{Neural networks},
  vol. 113, pp. 54--71, 2019.

\bibitem{mai2022online}
Z.~Mai, R.~Li, J.~Jeong, D.~Quispe, H.~Kim, and S.~Sanner, ``Online continual
  learning in image classification: An empirical survey,''
  \emph{Neurocomputing}, vol. 469, pp. 28--51, 2022.

\bibitem{liu2020learning}
B.~Liu, ``Learning on the job: Online lifelong and continual learning,'' in
  \emph{Proceedings of the AAAI Conference on Artificial Intelligence},
  vol.~34, no.~09, 2020, pp. 13\,544--13\,549.

\bibitem{BOX1979201}
\BIBentryALTinterwordspacing
G.~Box, ``Robustness in the strategy of scientific model building,'' in
  \emph{Robustness in Statistics}, R.~L. LAUNER and G.~N. WILKINSON, Eds.\hskip
  1em plus 0.5em minus 0.4em\relax Academic Press, 1979, pp. 201--236.
  [Online]. Available:
  \url{https://www.sciencedirect.com/science/article/pii/B9780124381506500182}
\BIBentrySTDinterwordspacing

\bibitem{arrieta2019search}
A.~Arrieta, S.~Wang, G.~Sagardui, and L.~Etxeberria, ``Search-based test case
  prioritization for simulation-based testing of cyber-physical system product
  lines,'' \emph{Journal of Systems and Software}, vol. 149, pp. 1--34, 2019.

\bibitem{panichella2018large}
A.~Panichella, F.~M. Kifetew, and P.~Tonella, ``A large scale empirical
  comparison of state-of-the-art search-based test case generators,''
  \emph{Information and Software Technology}, vol. 104, pp. 236--256, 2018.

\bibitem{silva2017systematic}
R.~A. Silva, S.~d. R.~S. de~Souza, and P.~S.~L. de~Souza, ``A systematic review
  on search based mutation testing,'' \emph{Information and Software
  Technology}, vol.~81, pp. 19--35, 2017.

\bibitem{henard2015combining}
C.~Henard, M.~Papadakis, M.~Harman, and Y.~Le~Traon, ``Combining
  multi-objective search and constraint solving for configuring large software
  product lines,'' in \emph{2015 IEEE/ACM 37th IEEE International Conference on
  Software Engineering}, vol.~1.\hskip 1em plus 0.5em minus 0.4em\relax IEEE,
  2015, pp. 517--528.

\bibitem{lopez2015systematic}
R.~E. Lopez-Herrejon, L.~Linsbauer, and A.~Egyed, ``A systematic mapping study
  of search-based software engineering for software product lines,''
  \emph{Information and software technology}, vol.~61, pp. 33--51, 2015.

\bibitem{zhang2020uncertainty}
H.~Zhang, M.~Zhang, T.~Yue, S.~Ali, and Y.~Li, ``Uncertainty-wise requirements
  prioritization with search,'' \emph{ACM Transactions on Software Engineering
  and Methodology (TOSEM)}, vol.~30, no.~1, pp. 1--54, 2020.

\bibitem{10.1007/978-3-540-69062-7_8}
Y.~Zhang, A.~Finkelstein, and M.~Harman, ``Search based requirements
  optimisation: Existing work and challenges,'' in \emph{Requirements
  Engineering: Foundation for Software Quality}, B.~Paech and C.~Rolland,
  Eds.\hskip 1em plus 0.5em minus 0.4em\relax Berlin, Heidelberg: Springer
  Berlin Heidelberg, 2008, pp. 88--94.

\bibitem{xu2006low}
Y.~Xu, X.~Cao, and H.~Qiao, ``A low-cost pedestrian detection system with a
  single optical camera,'' in \emph{2006 6th World Congress on Intelligent
  Control and Automation}, vol.~2.\hskip 1em plus 0.5em minus 0.4em\relax IEEE,
  2006, pp. 8759--8763.

\bibitem{rosin1997new}
C.~D. Rosin and R.~K. Belew, ``New methods for competitive coevolution,''
  \emph{Evolutionary computation}, vol.~5, no.~1, pp. 1--29, 1997.

\bibitem{8454482}
X.~Ma, X.~Li, Q.~Zhang, K.~Tang, Z.~Liang, W.~Xie, and Z.~Zhu, ``A survey on
  cooperative co-evolutionary algorithms,'' \emph{IEEE Transactions on
  Evolutionary Computation}, vol.~23, no.~3, pp. 421--441, 2019.

\bibitem{ouni2013maintainability}
A.~Ouni, M.~Kessentini, H.~Sahraoui, and M.~Boukadoum, ``Maintainability
  defects detection and correction: a multi-objective approach,''
  \emph{Automated Software Engineering}, vol.~20, pp. 47--79, 2013.

\bibitem{arcuri2014co}
A.~Arcuri and X.~Yao, ``Co-evolutionary automatic programming for software
  development,'' \emph{Information Sciences}, vol. 259, pp. 412--432, 2014.

\bibitem{wilkerson2012multi}
J.~L. Wilkerson, D.~R. Tauritz, and J.~M. Bridges, ``Multi-objective
  coevolutionary automated software correction,'' in \emph{Proceedings of the
  14th annual conference on Genetic and evolutionary computation}, 2012, pp.
  1229--1236.

\bibitem{koos2009automatic}
S.~Koos, J.-B. Mouret, and S.~Doncieux, ``Automatic system identification based
  on coevolution of models and tests,'' in \emph{2009 IEEE congress on
  evolutionary computation}.\hskip 1em plus 0.5em minus 0.4em\relax IEEE, 2009,
  pp. 560--567.

\bibitem{bollati2010environmental}
V.~Bollati and A.~Baccarelli, ``Environmental epigenetics,'' \emph{Heredity},
  vol. 105, no.~1, pp. 105--112, 2010.

\bibitem{liscovitch2017trade}
N.~Liscovitch-Brauer, S.~Alon, H.~T. Porath, B.~Elstein, R.~Unger, T.~Ziv,
  A.~Admon, E.~Y. Levanon, J.~J. Rosenthal, and E.~Eisenberg, ``Trade-off
  between transcriptome plasticity and genome evolution in cephalopods,''
  \emph{Cell}, vol. 169, no.~2, pp. 191--202, 2017.

\bibitem{vogt2017facilitation}
G.~Vogt, ``Facilitation of environmental adaptation and evolution by epigenetic
  phenotype variation: insights from clonal, invasive, polyploid, and
  domesticated animals,'' \emph{Environmental epigenetics}, vol.~3, no.~1,
  2017.

\bibitem{kelly2015churchill}
B.~J. Kelly, J.~R. Fitch, Y.~Hu, D.~J. Corsmeier, H.~Zhong, A.~N. Wetzel, R.~D.
  Nordquist, D.~L. Newsom, and P.~White, ``Churchill: an ultra-fast,
  deterministic, highly scalable and balanced parallelization strategy for the
  discovery of human genetic variation in clinical and population-scale
  genomics,'' \emph{Genome biology}, vol.~16, pp. 1--14, 2015.

\bibitem{kanherkar2014epigenetics}
R.~R. Kanherkar, N.~Bhatia-Dey, and A.~B. Csoka, ``Epigenetics across the human
  lifespan,'' \emph{Frontiers in cell and developmental biology}, vol.~2,
  p.~49, 2014.

\bibitem{stolfi2018epigenetic}
D.~H. Stolfi and E.~Alba, ``Epigenetic algorithms: A new way of building gas
  based on epigenetics,'' \emph{Information Sciences}, vol. 424, pp. 250--272,
  2018.

\bibitem{mukhlish2020reward}
F.~Mukhlish, J.~Page, and M.~Bain, ``Reward-based epigenetic learning algorithm
  for a decentralised multi-agent system,'' \emph{International Journal of
  Intelligent Unmanned Systems}, vol.~8, no.~3, pp. 201--224, 2020.

\end{thebibliography}

\end{document}